\documentclass[english,twocolumn,showpacs,preprintnumbers,amsmath,amssymb,prb,superscriptaddress]{revtex4}
\usepackage[T1]{fontenc}
\usepackage[latin9]{inputenc}
\usepackage{textcomp}
\usepackage{graphicx}
\usepackage{amssymb}

\makeatletter
\@ifundefined{textcolor}{}
{%
 \definecolor{BLACK}{gray}{0}
 \definecolor{WHITE}{gray}{1}
 \definecolor{RED}{rgb}{1,0,0}
 \definecolor{GREEN}{rgb}{0,1,0}
 \definecolor{BLUE}{rgb}{0,0,1}
 \definecolor{CYAN}{cmyk}{1,0,0,0}
 \definecolor{MAGENTA}{cmyk}{0,1,0,0}
 \definecolor{YELLOW}{cmyk}{0,0,1,0}
 }

\usepackage{color}

\makeatletter

\usepackage{placeins}

\usepackage{dcolumn}

\usepackage{bm}

\makeatother

\makeatother

\usepackage{babel}

\begin{document}

\title{Ballistic heat transport of quantum spin excitations as seen in SrCuO$_{2}$}

\author{N. Hlubek}

\author{P. Ribeiro}

\affiliation{IFW-Dresden, Institute for Solid State Research, P.O. Box 270116,
D-01171 Dresden, Germany}

\author{R. Saint-Martin}

\author{A. Revcolevschi}

\affiliation{Laboratoire de Physico-Chimie de L'Etat Solide, ICMMO, UMR8182, Université
Paris-Sud, 91405 Orsay, France}

\author{G. Roth}

\affiliation{Institut für Kristallographie der RWTH, D-52056 Aachen, Germany}

\author{G. Behr}

\author{B. Büchner}

\author{C. Hess}

\affiliation{IFW-Dresden, Institute for Solid State Research, P.O. Box 270116,
D-01171 Dresden, Germany}


\begin{abstract}
Fundamental conservation laws\textit{ }predict ballistic, i.e., \textit{dissipationless}
transport behaviour in one-dimensional quantum magnets. Experimental
evidence, however, for such anomalous transport has been lacking ever
since. Here we provide experimental evidence for ballistic heat transport
in a $S=1/2$ Heisenberg chain. In particular, we investigate high
purity samples of the chain cuprate $\mathrm{SrCuO_{2}}$ and observe
a huge magnetic heat conductivity $\kappa_{\mathrm{mag}}$. An extremely
large spinon mean free path of more than a micrometer demonstrates
that $\kappa_{\mathrm{mag}}$ is only limited by \textit{extrinsic}
scattering processes which is a clear signature of ballistic transport
in the underlying spin model. 
\end{abstract}
\maketitle
 The integrability of the one-dimensional (1D) antiferromagnetic
$S=1/2$ Heisenberg chain implies highly anomalous transport properties,
in particular, a \textit{divergent} magnetic heat conductivity $\kappa_{\mathrm{mag}}$
at all finite temperatures $T$.~\citep{Zotos1997,Zotos1999,Klumper2002,Heidrich2003,Heidrich2005}
This truly \textit{ballistic} heat transport suggests anomalously
large life times and mean free paths of the quantum spin excitations
and renders 1D quantum magnets intriguing candidates for spin transport
and quantum information processing.~\citep{Meier2003,Meier2003a,Santos2008}
However, despite the rigorous prediction, experimental evidence for
ballistic heat transport in quantum magnets is lacking. Nevertheless,
promising large $\kappa_{\mathrm{mag}}$ has been observed in a number
of cuprate compounds which realize 1D $S=1/2$ Heisenberg antiferromagnets~\citep{Hess2007,Sologubenko01,Hess2007b,Sologubenko00,Hess01,Hess04a,Kawamata2008,Hess06,Kudo01,Ribeiro05}
with the spin chain material $\mathrm{SrCuO_{2}}$ being a prominent
example~\citep{Sologubenko01,Ribeiro05} although a quantitative
analysis of $\kappa_{\mathrm{mag}}$ has always been difficult there
since the phononic and magnetic heat conductivities are of similar
magnitude at low temperature. Such experimental $\kappa_{\mathrm{mag}}$
is always \textit{finite} since extrinsic scattering processes due
to defects and phonons are inherent to all materials and mask the
intrinsic behavior of the chain. Formally it seems reasonable to account
for the extrinsic scattering via a finite $\kappa_{\mathrm{mag}}\sim D_{\mathrm{th}}\tau$,
where $\tau$ is a relaxation time, and $D_{\mathrm{th}}$ represents
the thermal Drude weight which (multiplied by a delta function at
zero frequency) describes the intrinsic heat conductivity. In fact,
it was thereby possible to identify the expected low-$T$ linearity
of $D_{\mathrm{th}}(T)$ in the case of a {}``dirty'' spin chain
material where a high density of chain defects generate a large $T$-independent
scattering rate $1/\tau$.~\citep{Hess2007}

In this paper we examine the heat conductivity of $\mathrm{SrCuO_{2}}$
which is considered an excellent realization of the $S=1/2$ Heisenberg
chain.~\citep{Motoyama1996,Matsuda97,Zaliznyak2004} Our samples
of extraordinary purity allow an unambiguous separation of the phononic
and magnetic contributions to the thermal conductivity. This yields
the by far highest $\kappa_{\mathrm{mag}}$ observed~\citep{Hess01,Sologubenko01}
until now. Our analysis reveals a remarkable lower bound for the low-$T$
limit of the mean free path $l_{\mathrm{mag}}$ of more than a micrometer.
Thus our data provide striking evidence that the intrinsic heat transport
of the S=1/2 Heisenberg chain is indeed ballistic. With increasing
temperatures $\kappa_{\mathrm{mag}}$ is increasingly supressed due
to spinon-phonon scattering which is the dominant extrinsic scattering
mechanism in this material. 

We have grown large single crystals of $\mathrm{SrCuO_{2}}$ by the
traveling solvent floating zone method,~\citep{Revcolevschi1999}
where the feed rods were prepared using the primary chemicals $\mathrm{CuO}$
and $\mathrm{SrCO_{3}}$ with both 2N (99\%) and 4N (99.99\%) purity.
Cuboidal samples with typical dimensions of $\left(3\times0.5\times0.5\right)\,\mbox{mm}^{3}$
were cut from the crystals, with the longest dimension parallel to
the principal axes. Four-probe measurements of the thermal conductivity
$\kappa$ were performed in the 7-300K range~\citep{Hess03a} with
the thermal current along the $a$, $b$, and $c$-axes ($\kappa_{a}$,
$\kappa_{b}$, and $\kappa_{c}$ respectively) for both the 2N and
the 4N samples. 

\begin{figure}
\begin{centering}
\includegraphics[bb=0bp 8bp 397bp 365bp,clip,width=1\columnwidth]{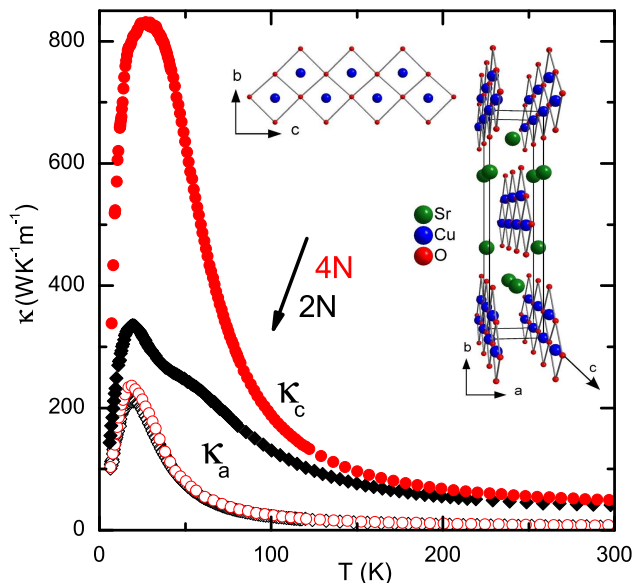}
\par\end{centering}

\caption{(Color online) $\kappa_{a}$ and $\kappa_{c}$ of SrCuO$_{2}$ for
different purity values. Closed (open) symbols represent $c$-axis
($a$-axis) data, circles (diamonds) correspond to 4N (2N) purity.
\label{fig:kappa}Inset: crystal structure of $\mathrm{SrCuO_{2}}$.
The symmetry is $Cmcm$ with lattice constants $a=3.56$~Å, $b=16.32$~Å,
$c=3.92$~Å.~\citep{Teske1970}}

\end{figure}

The main structural element in $\mathrm{SrCuO_{2}}$ is formed by
$\mathrm{CuO_{2}}$ zig-zag ribbons, which run along the crystallographic
$c$-axis (see inset Fig.~\ref{fig:kappa}). Each ribbon can be viewed
as made of two parallel corner-sharing $\mathrm{CuO_{2}}$ chains,
where the straight Cu-O-Cu bonds of each double-chain structure result
in a very large antiferromagnetic intrachain exchange coupling $J/k_{B}\approx2100-2600$
K of the $S=1/2$ spins at the $\mathrm{Cu^{2+}}$sites.~\citep{Zaliznyak2004,Motoyama1996}
The frustrated and much weaker interchain coupling $\left|J^{'}\right|/J\approx0.1-0.2$~\citep{Rice93,Motoyama1996}
and presumably quantum fluctuations prevent three-dimensional long
range magnetic order of the system at $T>T_{N}\approx1.5-2\,\mathrm{K}\approx10^{-3}J/k_{B}$~K.~\citep{Matsuda97,Zaliznyak1999}
Hence, at significantly higher $T$ the two chains within one double
chain structure can be regarded as magnetically independent. In fact,
low-$T$ (12 K) inelastic neutron scattering spectra of the magnetic
excitations can be very well described within the $S=1/2$ Heisenberg
antiferromagnetic chain model.~\citep{Zaliznyak2004}

Fig.~\ref{fig:kappa} presents our results for $\kappa_{a}$ and
$\kappa_{c}$ of SrCuO$_{2}$ for both 2N and 4N purity as a function
of $T$. We first describe the data for 2N purity which are in good
agreement with earlier results by Sologubenko et al.~\citep{Sologubenko01}
A pronounced low-$T$ peak at $\sim$18~K with $\kappa_{\mathrm{max}}\approx215\,\mathrm{Wm^{-1}K^{-1}}$
is found for $\kappa_{a,\mathrm{2N}}$, i.e., perpendicular to the
chains. This peak and a $\sim\!\! T^{-1}$-decrease at $T\gtrsim150$~K
towards a small value at room temperature ($\sim6\mathrm{\, Wm^{-1}K^{-1}}$)
represent the characteristic $T$-dependence of phonon-only heat conductivity
$\kappa_{\mathrm{ph}}$. The peak originates from two competing effects:~\citep{Berman}
at low $T$, a weakly $T$-dependent phonon mean free path $l_{\mathrm{ph}}$
and a rapidly increasing number of phonons cause $\kappa_{\mathrm{ph}}$
to increase strongly. At higher $T$, the exponentially rising number
of phonon-phonon umklapp processes increasingly shortens $l_{\mathrm{ph}}$,
which causes the decrease of $\kappa_{\mathrm{ph}}$. A similar low-$T$
peak (at $\sim20$~K) is also present in $\kappa_{c,\mathrm{2N}}$
(parallel to the chains). It is however larger ($\kappa_{\mathrm{max}}\approx335\mathrm{\, Wm^{-1}K^{-1}}$)
and exhibits a distinct shoulder at the high-$T$ edge ($T\gtrsim40$~K)
of the peak. $\kappa_{c,\mathrm{2N}}$ decreases at higher $T$, but
remains much larger than $\kappa_{a,\mathrm{2N}}$ and even at room
temperature $\kappa_{c,\mathrm{2N}}\approx40\,\mathrm{Wm^{-1}K^{-1}}$.
The apparent large anisotropy, together with the unusual $T$-dependence
of $\kappa_{c,\mathrm{2N}}$, is the signature of a large magnetic
fraction of $\kappa_{c,\mathrm{2N}}$ over a large $T$-range.~\citep{Sologubenko01,Ribeiro05}

We now turn to the new data which have been obtained for the high-purity
compound. The heat transport perpendicular to the chains ($\kappa_{a,\mathrm{4N}}$)
is slightly enhanced as compared to $\kappa_{a,\mathrm{2N}}$ ($\kappa_{\mathrm{max}}\approx235\mathrm{\, Wm^{-1}K^{-1}}$)
which reflects a somewhat reduced phonon-defect scattering. However,
a much more drastic and unexpected large effect of the enhanced purity
is observed in the heat transport parallel to the chains, $\kappa_{c,\mathrm{4N}}$.
Instead of a narrow low-$T$ peak and a shoulder as observed in $\kappa_{c,\mathrm{2N}}$,
a huge and broad peak centered at $\sim$28~K is present in $\kappa_{c,\mathrm{4N}}$
($\kappa_{\mathrm{max}}\approx830\mathrm{\, Wm^{-1}K^{-1}}$) which
exceeds $\kappa_{c,\mathrm{2N}}$ at $T\lesssim70$~K by more than
a factor of 2. Also at $70\,\mathrm{K}\lesssim T\leq300$~K we observe
$\kappa_{c,\mathrm{4N}}>\kappa_{c,\mathrm{2N}}$, where interestingly
both curves approach each other and at $T\gtrsim200$~K exhibit almost
the same $T$-dependence.

Without further analysis some clear-cut conclusions can be drawn.
First, the extraordinary enhancement of $\kappa_{c}$ upon the improvement
of the material's purity in contrast to a concomitantly negligibly
small one in $\kappa_{a}$, straightforwardly implies that the enhancement
primarily concerns the magnetic heat conductivity $\kappa_{\mathrm{mag}}$
which is present in $\kappa_{c}$ only. Second, the extreme low-$T$
sensitivity to impurities of $\kappa_{\mathrm{mag}}$ suggests that
spinon-defect scattering is the dominating process which relaxes the
heat current in this regime. Third, upon rising $T$, the spinon-defect
scattering is increasingly masked by a further scattering process
which leads to $\kappa_{c,\mathrm{2N}}$ and $\kappa_{c,\mathrm{4N}}$
being very similar at $T\gtrsim200$~K. The most reasonable candidate
for this process is spinon-phonon scattering, since the only thinkable
alternative, i.e. spinon-spinon scattering, is negligible~\citep{Hess2007,Klumper2002,Heidrich2003,Heidrich2005}
in this $T$-regime. 

\begin{figure}
\begin{centering}
\includegraphics[bb=0bp 8bp 397bp 365bp,clip,width=1\columnwidth]{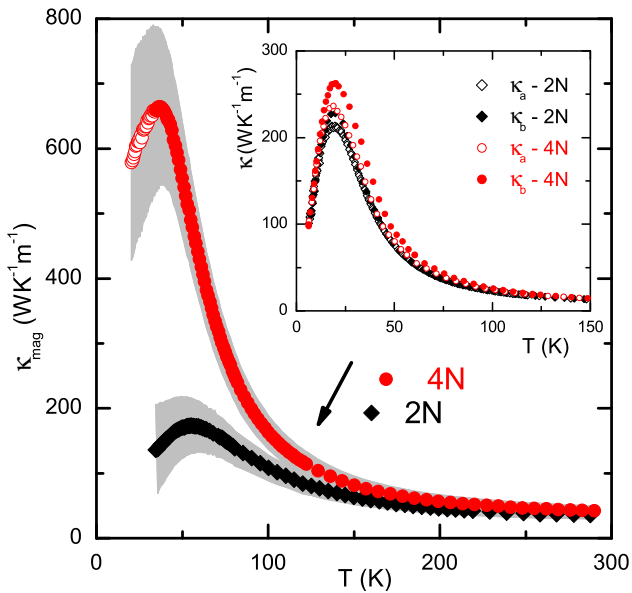}
\par\end{centering}

\caption{(Color online) $\kappa_{\mathrm{mag}}$ of SrCuO$_{2}$ for different
purities. Open symbols represent low-T $\kappa_{\mathrm{mag}}$ which
is disregarded in the further analysis. The shaded areas show the
uncertainty of the estimation of $\kappa_{\mathrm{mag}}$ due to the
phononic background. \label{fig:kappa_mag} Inset: $\kappa_{a}$ and
$\kappa_{b}$ perpendicular to the chain for both purities. }

\end{figure}

A further analysis of the data requires a reliable separation of the
total measured $\kappa$ into all relevant contributions which normally
add up. Since electronic contributions can be excluded in this electrically
insulating material, it seems natural to assume that the measured
$\kappa_{c}$ is just the sum of $\kappa_{\mathrm{mag}}$ and a phononic
background $\kappa_{\mathrm{ph},c}$,~\citep{Sologubenko00,Hess01,Sologubenko01,Hess04a,Hess2007,Hess2007b}
where the latter can be approximated by the purely phononic heat conductivity
perpendicular to the chains, $\kappa_{a}\approx\kappa_{b}$ (see inset
of Fig.~\ref{fig:kappa_mag}). The thus obtained $\kappa_{\mathrm{mag}}=\kappa_{c}-\kappa_{a}$
for the 2N and the 4N samples are shown in Fig.~\ref{fig:kappa_mag}.
At $T\lesssim35$~K, i.e., in the vicinity of the peak of $\kappa_{\mathrm{ph},c}$,
errors become large and we disregard the data in this range for further
analysis. For higher T we account for a possible uncertainty of $\pm30$\%
in $\kappa_{\mathrm{ph},c}$. Note that, in the case of the 4N compound,
possible errors in $\kappa_{\mathrm{mag}}$ are rendered small because
obviously $\kappa_{\mathrm{mag,4N}}\gg\kappa_{a,\mathrm{4N}}$. 

$\kappa_{\mathrm{mag}}$ of the 4N sample exhibits a sharp peak at
$\sim$37~K with an extraordinary maximum value of about $660\mathrm{\, Wm^{-1}K^{-1}}$,
which is more than a factor of 3 higher than the largest reported
$\kappa_{\mathrm{mag}}$.~\citep{Hess01,Sologubenko01} The peak
is followed by a strong decrease upon raising $T$. Similar to the
afore described typical $T$-dependence of a clean phononic heat conductor,
the overall $T$-dependence of $\kappa_{\mathrm{mag}}$ suggests that,
in a simple picture, two competing effects determine $\kappa_{\mathrm{mag}}$.
The low-$T$ increase of $\kappa_{\mathrm{mag}}$ is consistent with
a regime where the effect of scattering processes is weakly $T$-dependent
since $D_{\mathrm{th}}$ is expected to increase linearly with $T$.~\citep{Klumper2002,Hess2007,Heidrich2003,Heidrich2005}
The strong decrease at higher $T$ is then the result of the increasing
importance of spinon-phonon scattering. $\kappa_{\mathrm{mag}}$ of
the 2N sample is qualitatively very similar. However, the absolute
value at the peak is much lower ($\sim172$~W/mK) and the peak's
position is shifted to a higher $T$ ($\sim55$~K). Similarly to
the $\kappa_{c}$ data, at higher $T$, the 4N curve approaches that
of the 2N sample. The latter is consistent with the earlier notion
that spinon-phonon scattering is dominant at high $T$, while the
differences at low $T$ suggest that spinon-phonon scattering freezes
out, upon decreasing $T$, rendering spinon-defect scattering increasingly
important.

\begin{figure}
\begin{centering}
\includegraphics[bb=0bp 8bp 397bp 335bp,clip,width=1\columnwidth]{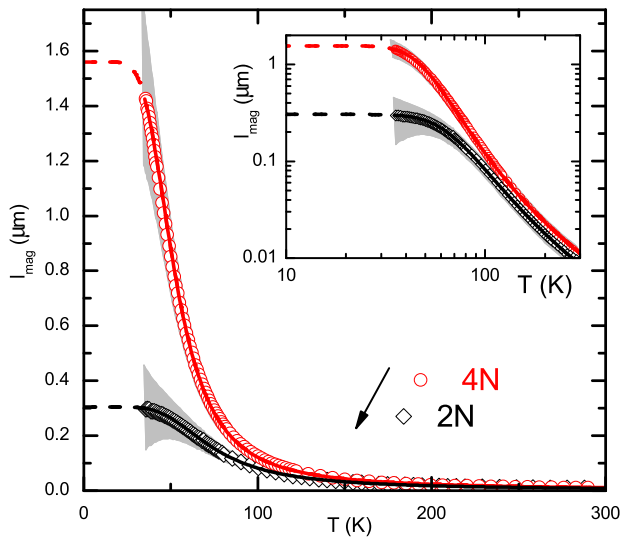}
\par\end{centering}

\caption{(color online) Magnetic mean free paths of SrCuO$_{2}$ for different
purities. The solid lines were calculated according to Eq.~\ref{eq:Umklapp}.
The shaded area illustrates the uncertainty from the estimation of
the phononic background. \label{fig:lmag}}

\end{figure}

We analyze $\kappa_{\mathrm{mag}}$ quantitatively by extracting the
spinon mean free path $l_{\mathrm{mag}}$ according to~\citep{Hess2007,Sologubenko01,Hess2007b}
\begin{equation}
l_{\mathrm{mag}}=\frac{3\hbar}{\pi N_{s}k_{B}^{2}T}\kappa_{\mathrm{mag}},\end{equation}
where $N_{s}=4/ab$ is the number of spin chains per unit area. As
can be inferred from Fig.~\ref{fig:lmag}, $l_{\mathrm{mag}}$ of
both samples show a strong decrease with increasing $T$, which directly
reflects spinon-phonon scattering becoming increasingly important.
Both curves are very similar, but clear differences are present at
low $T$, where $l_{\mathrm{mag}}$ of the 2N sample is somewhat lower,
in accordance with a higher spinon-defect scattering. We evoke Matthiesen's
rule to model the $T$-dependence of $l_{\mathrm{mag}}$ and to account
for both scattering processes viz. $l_{\mathrm{mag}}^{-1}=l_{0}^{-1}+l_{\mathrm{sp}}^{-1}$.
Here, $l_{0}$ describes the $T$-independent spinon-defect scattering
whereas $l_{\mathrm{sp}}(T)$ accounts for the $T$-dependent spinon-phonon
scattering. For the latter, we assume a general umklapp process with
a characteristic energy scale $k_{B}T_{u}^{*}$ of the order of the
Debye energy, which is commonly used in literature.~\citep{Sologubenko01,Kawamata2008}
We thus have\begin{equation}
l_{\mathrm{mag}}^{-1}=l_{0}^{-1}+\left(\frac{\exp\left(T_{u}^{*}/T\right)}{A_{s}T}\right)^{-1},\label{eq:Umklapp}\end{equation}
which can be used to fit the data with $l_{0}$, $A_{s}$ and $T_{u}^{*}$
($A_{s}$ describes the coupling strength) as free parameters. We
find an excellent agreement between such fits and the experimental
$l_{\mathrm{mag}}$, see Fig.~\ref{fig:lmag}. Inspection of the
fit parameters %
\footnote{For the 4N and 2N compounds we get $l_{0,4N}=\left(1.56\pm0.16\right)\mbox{\textmu m}$,
$T_{u,4N}^{*}=\left(204\pm11\right)\mbox{K}$, $A_{s,4N}=\left(58.6\pm5.4\right)$~$10^{-16}$m/K
and $l_{0,2N}=\left(305\pm5\right)\mbox{nm}$, $T_{u,2N}^{*}=\left(217\pm3\right)\mbox{K}$,
$A_{S,2N}=\left(78\pm1\right)$~$10^{-16}$m/K, respectively. Setting
$T_{u,2N}^{*}=T_{u,4N}^{*}=204$~K after fitting the 4N data gives
$l_{0,2N}=\left(320\pm13\right)\mbox{nm}$ and $A_{s,2N}=\left(72\pm5\right)$~$10^{-16}$m/K
for 2N. The errors account for the accuracy of the fit. It is also
possible to obtain a good fit with the same $A_{s}$ for the 4N and
2N cases. However, individual $A_{s}$ account for errors in the absolute
value, while fixing the energy scale by $T_{u}^{*}$ seems physically
justified. %
} yields two remarkable aspects which corroborate our previous qualitative
findings. First, the parameters $A_{s}$ and $T_{u}^{*}$ which determine
the spinon-phonon scattering are practically the same for both samples.
In fact, an equally good fit is obtained if the \textit{same }$T_{u}^{*}$
is used for both curves. Note that the extracted $T_{u}^{*}\sim200$~K
is indeed of the order of the Debye temperature $\Theta_{D}$ of this
material and thus leads to the conjecture that mostly acoustic phonons
are involved in this scattering process. %
\footnote{Modeling $l_{\mathrm{mag}}$ with $l_{\mathrm{sp}}\propto\exp(T^{*}/T)$
which accounts for the alternative scenario of spinons scattering
off optical phonons~\citep{Hess05} results in a fit of similar quality
with comparable $l_{0}$ and $T^{*}\sim300$~K.%
} Second, the spinon-defect scattering length $l_{0}$, which represents
a lower bound for the low-$T$ limit of $l_{\mathrm{mag}}$ and which
should significantly depend on the sample's purity turns out to be
drastically different for both cases. To be specific, we find $l_{0}\approx300$~nm
for the 2N compound and an extraordinary $l_{0}\approx1.6$~\textmu{}m
for the 4N sample, which correspond to more than 750 and 4100 lattice
spacings, respectively. These findings provide a further confirmation
of the above interpretation that, in both cases, $\kappa_{\mathrm{mag}}$
is determined by the same spinon-phonon scattering process and that
the difference between the two curves can be described by the different
defect density only. We mention that our results are consistent with
recent data by T. Kawamata et al.~ %
\footnote{T. Kawamata, N. Kaneko, M. Uesaka, M. Sato, and Y. Koike, unpublished
data.%
}

A major outcome of our study is the unambiguous identification of
the \textit{extrinsic} scattering processes as the only relevant ones.
\textit{Intrinsic} spinon-spinon scattering, on the other hand, plays
no role in our analysis, even in the case of the very clean sample.
The strong enhancement of $\kappa_{\mathrm{mag}}$ upon reduction
of the impurity amount thus appears as the manifestation of ballistic
heat transport of the underlying spin model, where $\kappa_{\mathrm{mag}}$
is rendered finite by extrinsic scattering processes only. One might
therefore speculate that $\kappa_{\mathrm{mag}}$ of this material
can be driven to much higher values in a perfect crystal. 

We point out that our analysis relies on a very simple theoretical
approach which was also successfully used in many other low-dimensional
$S=1/2$ spin systems,~\citep{Hess01,Hess03,Hess04a,Hess05,Hess06,Hess2007,Sologubenko00,Sologubenko01,Kawamata2008,Hess2007b}
which is surprising in view of the strong quantum nature of such systems.
More sophisticated approaches might lead to a deeper understanding
of the magnetic heat transport in this system on a microscopic level.
In this regard it is interesting to note that, in clean samples (i.e.
with large $l_{0}$), $l_{\mathrm{sp}}=\left(A_{s}T\right)^{-1}\exp(T_{u}^{*}/T)$
leads to $\kappa_{c}\approx\kappa_{\mathrm{mag}}\propto\exp(T_{u}^{*}/T)$
with $T_{u}^{*}\sim200$~K at high $T$, in agreement with the theory
proposed by Shimshoni et al.~\citep{Shimshoni03} However, we do
not observe $\kappa_{a}=\kappa_{\mathrm{ph}}\propto\exp(2T_{u}^{*}/T)$
as expected in the same model. It seems worthwhile mentioning in this
regard that the only slight enhancement of $\kappa_{a}$ observed
upon increasing purity is quite unexpected. One might speculate that
this is an indication of phonon scattering off the spin chains which
in principle should be relevant.~\citep{Chernyshev05,Rozhkov05}

To sum up, we have investigated the spinon heat conductivity $\kappa_{\mathrm{mag}}$
of the antiferromagnetic S=1/2 Heisenberg chain cuprate $\mathrm{SrCuO_{2}}$
for standard (99\%) and high (99.99\%) purity. The higher purity leads
to a drastic enhancement of $\kappa_{\mathrm{mag}}$ at low $T$ and
we find the up-to-present by far highest reported $\kappa_{\mathrm{mag}}$
in the high-purity sample. For higher $T$, we provide clear-cut evidence
that spinon-phonon scattering is the most relevant scattering which
leads to a very efficient reduction of $\kappa_{\mathrm{mag}}$. An
extreme sensitivity of $\kappa_{\mathrm{mag}}$ to impurities is present
at low $T$, which implies that the spinon-defect scattering is dominating
in this regime. A simple analysis reveals a remarkable lower bound
for the low-temperature limit of the spinon mean free path $l_{\mathrm{mag}}$
of more than a micrometer. Our results therefore suggest that $\kappa_{\mathrm{mag}}$
is only limited by extrinsic scattering processes which appears as
the manifestation of the ballistic nature of heat transport in the
$S=1/2$ antiferromagnetic Heisenberg chain. 

\begin{acknowledgments}

We thank W. Brenig, A. L. Chernyshev, S.-L. Drechsler, F. Heidrich-Meisner,
P. Prelovšek, X. Zotos and A. A. Zvyagin for stimulating discussions.
This work was supported by the Deutsche Forschungsgemeinschaft through
grant HE3439/7, through the Forschergruppe FOR912 (grant HE3439/8)
and by the European Commission through the NOVMAG project (FP6-032980).

\end{acknowledgments}

\bibliographystyle{apsrev}

\end{document}